\documentclass[12pt,aps,prb,preprint]{revtex4}

\usepackage{amsmath}
\usepackage{graphicx}

\begin{document}

\title{Gravity assist in 3D like in Ulysses mission}
\author{Herbert Morales}
 \affiliation{Escuela de F\'isica, Universidad de Costa Rica,
  San Jos\'e, Costa Rica}
 \email{hmorales@fisica.ucr.ac.cr}
\date{\today}

\begin{abstract}
We study the gravity assist in the general case, i.e.~when the
spacecraft is not in a coplanar motion with respect to the planet's
orbit.
Our derivation is based on Kepler's planetary motion and
Galilean addition of velocities, subjects covered in introductory
physics courses.
The main purpose of this paper is to illustrate how the gravity
assist can be used to deviate a spacecraft outside its original
plane of motion.
As an example, we use the NASA-ESA's Ulysses mission to ``test"
our simple model.
\end{abstract}

\maketitle


\section{Introduction}

It is well known that gravity assist is an excellent technique to
speed up or slow down a spacecraft, but is also a mechanism to
deviate its course.
For these reasons, NASA and ESA use the gravity assist as a helpful
tool to reduce mission cost and to save fuel and travel time.
Otherwise, planet exploration would be hard to pursue
using current technologies.

This phenomenon, also called slingshot effect or swing-by maneuver,
can be easily understood as an elastic collision of two particles.
\cite{BH85, V03, G04}
However, there is a paradox in the elastic derivation, because
if one assumes that the planet's mass is much greater than the
spacecraft's, one would conclude that the initial and final speeds
of the spacecraft (both when it is very far from the planet) should be
equal, like in the usual Kepler motion.
The ``mistake" with this thinking is that one is forgetting the
reference frame, since the above conclusion is correct for an
observer on the planet, but not for an observer on the Sun due to
the relative motion between the planet and the Sun.
Therefore, one needs to make a Galilean addition of velocities
for the latter case that results in the swing-by maneuver.
One can check that the elastic collision is satisfied, the
energy gained by the spacecraft is equal to the energy lost by the
planet, but this is so small that the planet's motion is not
altered. \cite{V03}
The reader can also look for other approaches to explain
the slingshot effect, like derivations from special relativity,
\cite{DCG04, CDG05}
classical Lagrangian \cite{E05} and work-impulse. \cite{C05}
Our goal in this paper is to get some insights of how the gravity
assist changes the course of a spacecraft away from the ecliptic
plane like in Ulysses mission.
Our motivation is that we can find no literature about
3D gravity assist at the introductory physics level. 

Let us start with the description of our assumptions.
We will use the word {\it planet} for the celestial body that
alters the spacecraft's trajectory.

\begin{enumerate}

\item For simplicity, the Sun, the Earth and the planet are all in the
same plane, the {\it ecliptic plane.}
We also consider that the Sun's and the planet's equators
lie in that plane.
\label{ecliptica}

\item The spacecraft motion can be divided in four parts,
according to which celestial object contributes more to the
gravitational force:
\begin{enumerate}
\item escaping the Earth,
\item moving by Sun interaction,
\item ``swinging by" the planet and
\item orbiting around the Sun or escaping the Solar System.
\end{enumerate}
We concentrate our work on the last two parts, where
the gravity assist effect is employed and causes the final
trajectory.
\label{motions}

\item After escaping the Earth, the spacecraft moves in a
plane that is parallel to ecliptic.

\item The interaction between the planet and the spacecraft is
considered as scattering phenomenon in the planet's frame,
occurring in a very short time compared to the period of the
planet's circular orbit.
Therefore, the planet's motion is taken as a straight line during
the slingshot effect.
\label{scatter}

\end{enumerate}


\section{Hyperbolic Trajectory Review \label{sec:hyp}}

\begin{figure}[h]
\begin{center}
\scalebox{0.80}{\includegraphics{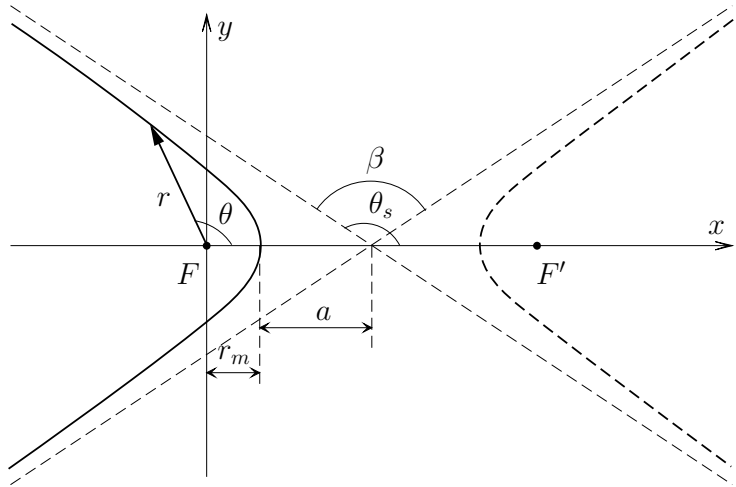}}
\caption{Hyperbolic trajectory.
$F$ is the focal point where the force center is located.}
\label{fig:hyper}
\end{center}
\end{figure}

Our goal in this section is to determine the scattering angle of the 
spacecraft, $\beta$, due to the gravitational interaction with the
planet, in the planet's frame.
A hyperbolic trajectory is shown in Fig.~\ref{fig:hyper}.
In polar coordinates, the equation for this kind of conic section is
given by
\begin{equation}
r = \frac{a\, (\varepsilon^2 - 1)}{1 + \varepsilon \cos \theta}\, ,
\label{eq:cse}
\end{equation}
where $\varepsilon ( > 1)$ is the eccentricity and $a$ is the
semi-major axis.

As seen from this equation, a particular hyperbola is completely
defined by two parameters: $\varepsilon$ and $a$.
In celestial mechanics, these parameters are usually written as functions
of the first integrals of the motion: the energy, $E$, and the angular
momentum, $l$, of a particle moving under the influence of a central
inverse-square law force.
In scattering problems, the impact parameter is used in place of the
angular momentum.

We consider that the initial velocity $u_i$ and the pericenter $r_m$
are our known values, so that they will take the place of the above
parameters,	 describing the spacecraft's hyperbolic orbit.

By solving for the eccentricity in Eq.~(\ref{eq:cse}) when $\theta = 0$
and $ r = r_m$, we find
\begin{equation}
\varepsilon = 1 + \frac{r_m}{a}\, ,
\end{equation}
where the semi-major axis is obtained by the usual celestial-mechanics
equation
\begin{equation}
a \equiv \frac{G M m}{2 E} = \frac{G M}{u_i^2}\, ,
\label{major}
\end{equation}
since $E = \frac{1}{2} m u_i^2$, when the spacecraft is far away from
central force source.

By taking $r \to \infty$ in the conic section equation (\ref{eq:cse})
and solving for $\theta$, we determine the angle between the asymptote
and the $x$-axis
\begin{equation}
\cos \theta_s = - \frac{1}{\varepsilon}\, .
\end{equation}

With the geometry shown in Fig.~\ref{fig:hyper}, the spacecraft's
scattering angle is
\begin{equation}
\beta = 2 \theta_s - \pi,
\end{equation}
which has values $0 \le \beta \le \pi$ (attractive scattering).


\section{The simple 3D case for gravity assist \label{toy}}

\begin{figure}[h]
\begin{center}
\scalebox{0.8}{\includegraphics{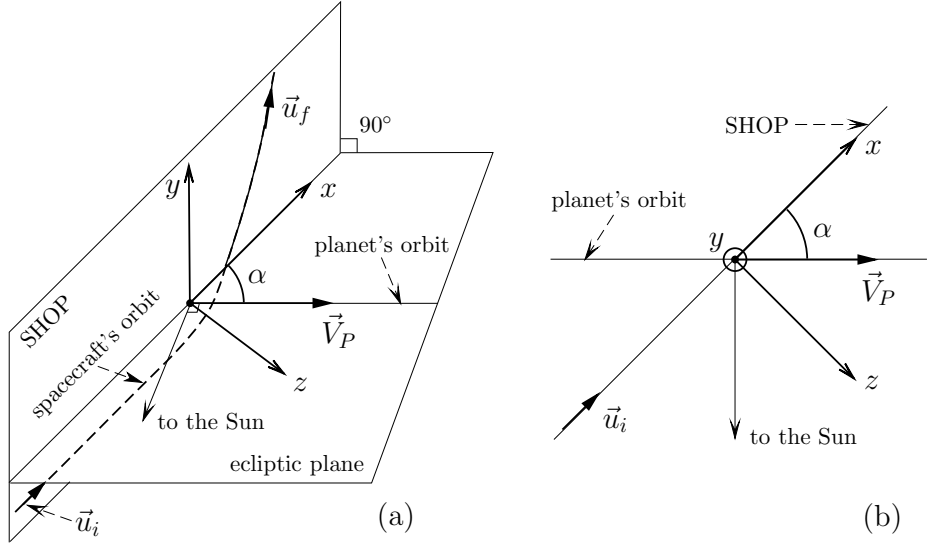}}
\caption{The simple 3D case for gravity assist:
(a) its visual representation in 2D and (b) the top view.}
\label{fig:compVp}
\end{center}
\end{figure}

Let us first study the case when the spacecraft's-hyperbolic-orbit
plane (SHOP) is perpendicular to the ecliptic plane
(Fig.~\ref{fig:compVp}).
Note that under assumption \ref{scatter}, SHOP is a fixed plane in
the planet's frame, but it is a constant-velocity moving plane in the
Sun's frame.
For this particular case, the 3D vector analysis gets simplified and
allows us to picture how the general case can be established.
We also choose that the spacecraft flies over the planet's
south pole.
Our goal here is to determine the angle between the
spacecraft's-final-orbit plane (SFOP) and the ecliptic plane,
we will call it the {\it elevation angle}, $\gamma$.

In a planet's frame as that shown in Fig.~\ref{fig:compVp}, the
components of the planet's velocity relative to the Sun,
$\vec{V}_P$, are
\begin{equation}
V_{P,x} = V_P \cos \alpha,
\quad
V_{P,y} = 0,
\quad
V_{P,z} = V_P \sin \alpha,
\end{equation}
where $\alpha$ is the angle between the planet's velocity
and the intersection of SHOP and the ecliptic plane.
We consider $ 0 \le \alpha \le \pi$, meaning that the spacecraft
is approaching a planet outside the Earth's orbit.

The initial and final velocities of the spacecraft in the Sun's frame,
$\vec{v}_i$ and $\vec{v}_f$, are obtained by Galilean addition
of velocities
\begin{equation}
\vec{v}_i = \vec{u}_i + \vec{V}_P,\quad
\vec{v}_f = \vec{u}_f + \vec{V}_P,
\label{galileo}
\end{equation}
where $\vec{V}_P$ obviously has the role of the velocity of the
planet's frame with respect to the Sun's frame and $\vec{u}_i$ and
$\vec{u}_f$ are the spacecraft's initial and final velocities in
the planet's frame.

\begin{figure}[h]
\begin{center}
\scalebox{0.8}{\includegraphics{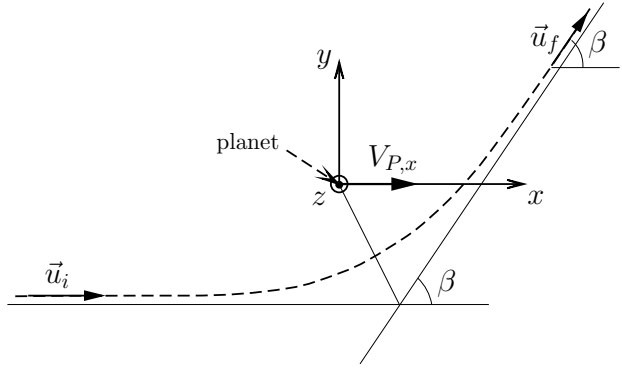}}
\caption{The spacecraft's-hyperbolic-orbit plane (SHOP) and the
spacecraft's motion as seen from the planet.
The vector $\vec{V}_P$ lies in the $zx$-plane.}
\label{fig:VSun}
\end{center}
\end{figure}

With the aid of Fig.~\ref{fig:VSun}, we find the magnitude $v_i$
to be
\begin{equation}
v_i^2 = (u_i + V_{P,x})^2 + V_{P, y}^2 + V_{P, z}^2
      = u_i^2 + V_P^2 + 2 u_i V_P \cos \alpha.
\label{eq:Vinitial}
\end{equation}
Note that $\alpha$ is also the angle between $\vec{V}_P$ and
$\vec{u}_i$.

For the final velocity $\vec{v}_f$, we first compute the components
of $\vec{u}_f$ in the planet's frame
\begin{equation}
u_{f,x} = u_f \cos \beta,
\quad
u_{f,y} = u_f \sin \beta.
\label{eq:uf}
\end{equation}
Therefore, the magnitude $v_f$ is given by 
\begin{equation}
v_f^2 = (u_{f,x} + V_{P,x})^2 + (u_{f,y} + V_{P,y})^2 + V_{P,z}^2
      = u_f^2 + V_P^2 + 2 u_f V_P \cos \alpha\, \cos \beta.
\label{eq:Vfinal}
\end{equation}

Then, we subtract Eq.~(\ref{eq:Vinitial}) from Eq.~(\ref{eq:Vfinal})
and obtain
\begin{equation}
v_f^2 - v_i^2 = 2 u V_P \cos \alpha\, (\cos \beta - 1),
\label{eq:VfVi}
\end{equation}
since $u_i = u_f = u$ by assumption \ref{scatter}.

Eq.~(\ref{eq:VfVi}) allows us to distinguish three situations according
to how initial and final velocities in the Sun's frame are related,
\begin{itemize}
\item $v_f > v_i$: The spacecraft increases its speed, if
the planet scatters it ($\beta \ne 0$) and
they encounter each other ($\frac{\pi}{2} < \alpha \le \pi$),
\item $v_f < v_i$: It slows down, if
the planet scatters it ($\beta \ne 0$) and
the spacecraft ``tries" to catch the planet
($0 \le \alpha < \frac{\pi}{2}$),
\item $v_f = v_i$: There is no speed change, if
there is no scattering ($\beta = 0$) or
the spacecraft moves exactly perpendicular to the planet's motion
($\alpha = \frac{\pi}{2}$).
\end{itemize}

\begin{figure}[h]
\begin{center}
\scalebox{0.9}{\includegraphics{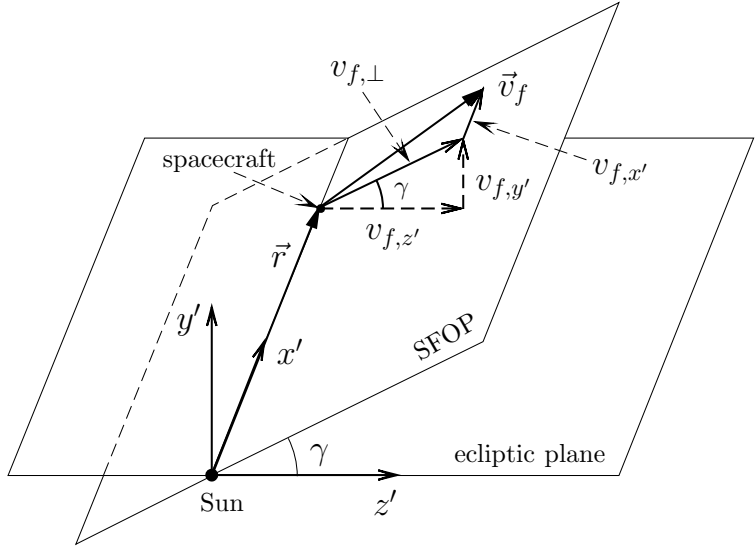}}
\caption{The spacecraft's-final-orbit plane (SFOP) in the Sun's frame.
$v_{f,\perp}$ is like the lever arm.}
\label{fig:SFOP}
\end{center}
\end{figure}

After the slingshot effect has occurred, we can determine the
spacecraft's final orbit.
According to assumption \ref{motions}, the initial kinematic
variables for the spacecraft orbiting around the Sun [part (d)] are
those at the end of the slingshot [part (c)].
Thus, the initial velocity in part (d) is equal to $\vec{v}_f$ and
the initial vector position $\vec{r}$ is taken to have a magnitude
equal to the planet's distance from the Sun and lying in the ecliptic
plane (Fig.~\ref{fig:SFOP}).
This approximation can be justified because the gravitational force
due to the planet is much weaker than that due to the Sun (see a
numerical calculation for Jupiter at the end of
Sec.~\ref{sec:numerical}).

From the angular momentum conservation and the properties of
the cross product, the SFOP is completely defined by $\vec{v}_f$
and $\vec{r}$.
Therefore, we conclude through Fig.~\ref{fig:SFOP} that
\begin{equation}
\tan \gamma = \frac{v_{f,y'}}{v_{f,z'}}\, ,
\label{eq:elevation}
\end{equation}
where these velocity components are defined in the Sun's frame  
that has the $x$-axis aligned in the direction of the vector $\vec{r}$.

\begin{figure}[h]
\begin{center}
\includegraphics{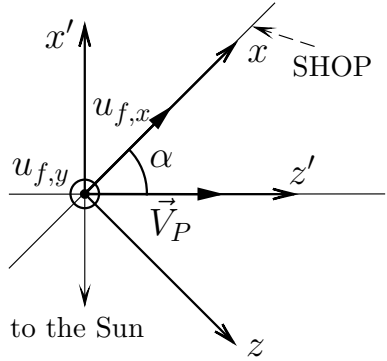}
\caption{Top view for the velocity components of the spacecraft
in the Sun's frame.}
\label{fig:angle}
\end{center}
\end{figure}

We recall Eq.~(\ref{galileo}) and use Fig.~\ref{fig:angle} to find
\begin{equation}
\tan \gamma = \frac{u_{f,y}}{V_P + u_{f,x} \cos \alpha}
    = \frac{\sin \beta}{(V_P/u) + \cos \alpha\, \cos \beta}\, .
\end{equation}
Note that we have derived an expression that depends only on
information before the slingshot.

\section{The general 3D case for gravity assist}

Let us now study the case when the angle between SHOP and the
ecliptic plane is some value $\delta$, not necessarily $90^\circ$.
We repeat the same derivation procedure done in the above section,
but including now the effect of the angle $\delta$.

\begin{figure}[h]
\begin{center}
\includegraphics{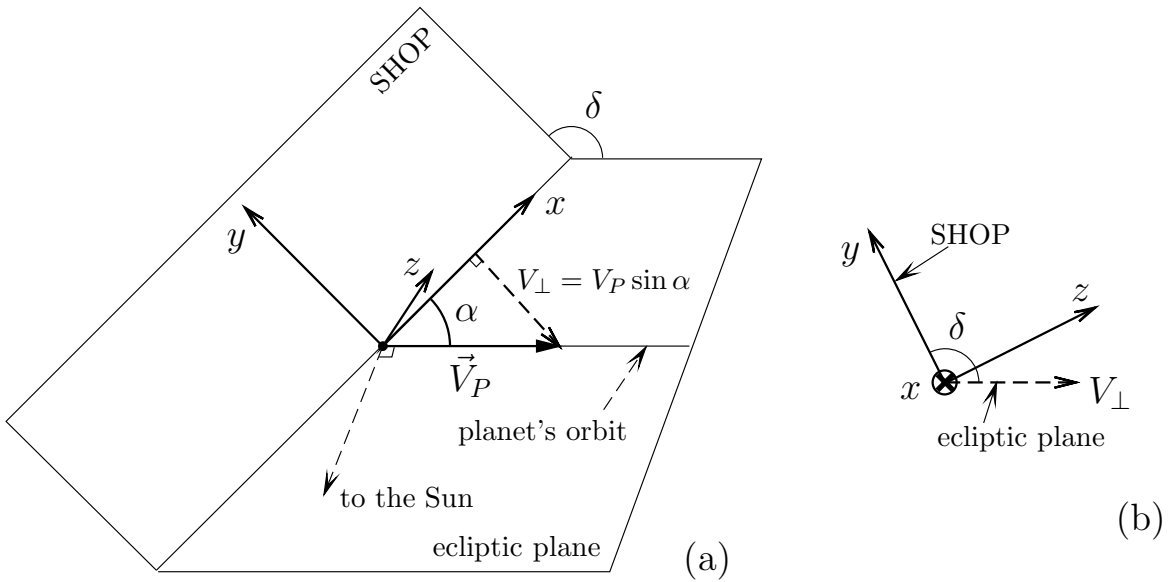}
\caption{The general slingshot case:
(a) its visual representation in 2D and (b) the front view.}
\label{fig:SOP-EP}
\end{center}
\end{figure}

From Fig.~\ref{fig:SOP-EP}, the components of $\vec{V}_P$ in the
planet's frame are
\begin{equation}
V_{P,x} = V_P \cos \alpha,
\quad
V_{P,y} = V_P \sin \alpha\, \cos \delta ,
\quad
V_{P,z} = V_P \sin \alpha\, \sin \delta,
\label{eq:Vp}
\end{equation}
where $\alpha$ is still the angle between the planet's velocity
and the intersection of SHOP and the ecliptic plane.

Using Fig.~\ref{fig:VSun} and Eq.~(\ref{galileo}), we find the
magnitude $v_i$ as we did before
\begin{equation}
v_i^2 = u_i^2 + V_P^2 + 2 u_i V_P \cos \alpha.
\label{eq:vi}
\end{equation}
Note that $\alpha$ still represents the angle between $\vec{V}_P$ and
$\vec{u}_i$ and the velocity components of the spacecraft in the
planet's frame, $\vec{u}_i$ and $\vec{u}_f$, do not change due to the
SHOP orientation.

The magnitude $v_f$ follows from Eqs.~(\ref{eq:uf}) and (\ref{eq:Vp})
in Eq.~(\ref{galileo})
\begin{equation}
v_f^2 = u_f^2 + V_P^2 + 2 u_f V_P (\cos \alpha\, \cos \beta +
        \sin \alpha\, \sin \beta\, \cos \delta).
\label{eq:vf}
\end{equation}

Then, we can subtract Eq.~(\ref{eq:vi}) from Eq.~(\ref{eq:vf}) and
obtain
\begin{equation}
v_f^2 - v_i^2 = 2 u V_P \left(\sin \alpha\, \sin \beta\, \cos \delta +
\cos \alpha\, (\cos \beta - 1) \right),
\end{equation}
since $u_i = u_f = u$.
With this result, the reader can study the ranges of $\alpha$,
$\beta$ and $\delta$ that make the spacecraft increase or
decrease its speed or even those that lead to no speed change.
She can also verify the usual 2D case ($\delta = 0, \pi$).
\cite{BH85, V03, G04}

\begin{figure}[h]
\begin{center}
\includegraphics{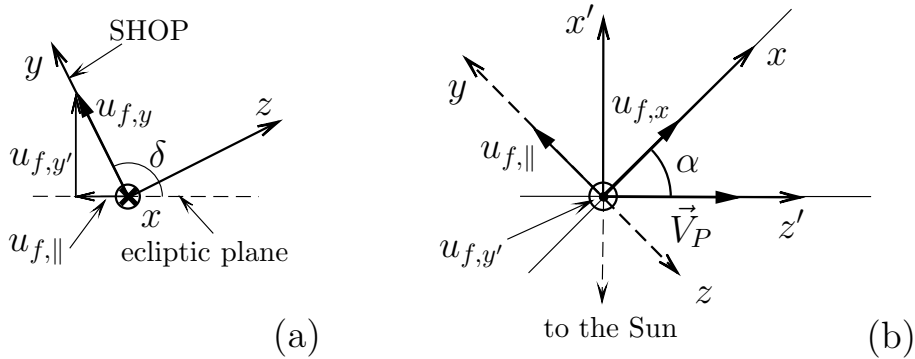}
\caption{The velocity components in the Sun's frame:
(a) front view and (b) top view.}
\label{fig:angle-grl}
\end{center}
\end{figure}

From the Fig.~\ref{fig:angle-grl} and Eq.~(\ref{eq:elevation}),
the elevation angle is given by
\begin{eqnarray}
\tan \gamma = \frac{v_{f,y'}}{v_{f,z'}}
   & = & \frac{u_{f,y} \sin \delta}{V_P + u_{f,x} \cos \alpha -
u_{f,y} \cos \delta\, \cos (\pi/2 + \alpha)}
\nonumber \\
    & = & \frac{\sin \beta\, \sin \delta}{(V_P/u) + \cos \alpha\,
\cos \beta + \sin \alpha\, \sin \beta\, \cos \delta}\, .
\label{eq:angle-grl}
\end{eqnarray}


\section{Numerical calculations \label{sec:numerical}}

Let us compute the elevation angle using Ulysses spacecraft data.
From any college physics textbook,\cite{SJ8} we have the following
Jupiter information:
\begin{equation*}
M_J = 1.90 \times 10^{27}\ {\rm kg},
\quad
a_J = 7.78 \times 10^{11}\ {\rm m},
\quad
R_J = 6.99 \times 10^7\ {\rm m},
\quad
\tau_J = 3.74 \times 10^8\ {\rm s},
\end{equation*}
where they are its mass, its distance from the Sun, its radius and
its period, respectively.

From Ulysses web page,\cite{JPL} we have the spacecraft information:
\begin{equation*}
m = 366.7\ {\rm kg},
\quad
u = u_i = 13.896\ {\rm km/s},
\quad
v_i = 16.184\ {\rm km/s},
\quad
r_m = 6.3\, R_J.
\end{equation*}
Therefore, the equations of Sec.~\ref{sec:hyp} result in
($G = 6.67 \times 10^{-11}\ \rm N \cdot m^2/kg^2$)
\begin{equation*}
a = 6.56 \times 10^8\ {\rm m},
\quad
\varepsilon = 1.67,
\quad
\theta_s = 127^\circ,
\quad
\beta = 74^\circ.
\end{equation*}
Assuming uniform circular motion, we estimate Jupiter's speed, $V_P$,
and use Eq.~(\ref{eq:vi}) to get $\alpha$,
\begin{equation*}
V_P = \left( \frac{2 \pi}{\tau_J} \right) a_J = 13.1\ {\rm km/s},
\quad
\alpha = 106^\circ.
\end{equation*}

We let the angle $\delta$ be a free parameter.
Table \ref{tab:elevation} shows the final speed and the elevation angle
computed from Eqs.~(\ref{eq:vf}) and (\ref{eq:angle-grl}) with
the above data.
\begin{table}[h]
\begin{center}
\begin{tabular}{|c|c|c|}
\hline
$\delta$ (deg) & $v_f\ \rm (km/s)$ & $\gamma$ (deg)
\\
\hline
0 & 26.0 & 0.0 \\
15 & 25.7 & 8.0 \\
30 & 25.1 & 16.1 \\
45 & 24.0 & 24.1 \\
60 & 22.5 & 32.1 \\
90 & 18.4 & 48.0 \\
120 & 13.0 & 64.1 \\
146.9 & 7.4 & 80.0 \\
150 & 6.8 & 82.1 \\
159.7 & 4.6 & 90.0 \\
165 & 3.5 & 95.9 \\
170 & 2.4 & 104.5 \\
175 & 1.4 & 122.7 \\
180 & 0.8 & 180.0 \\
\hline
\end{tabular}
\caption{The final speed and the elevation angle as functions of
$\delta$.}
\label{tab:elevation}
\end{center}
\end{table}
The usual condition for elliptic orbit $E < 0$ can be rewritten
as $v < \sqrt{2 G M_S / r} = 18.5$ km/s, with $r = a_j$ as it was
considered at the end of Sec.~\ref{toy} for deriving $\gamma$
($M_S = 1.99 \times 10^{30}$ kg).

From Table \ref{tab:elevation}, we conclude that the minimum elevation
angle that would keep the spacecraft orbiting around the Sun is
around $48^\circ$, meaning that the spacecraft would fly over a
Jupiter's pole.
Lower values of $\gamma$ would imply that the spacecraft will
escape the Solar System.
Notice that $\delta$ represents the maximum latitude (both north and
south) that SHOP crosses in the planet, under the assumption
\ref{ecliptica} (for $\delta > 90^\circ$, the related parallel is
$180^\circ - \delta$).
By symmetry, one could infer the results for
$180^\circ < \delta < 360^\circ$.
When the spacecraft encounters the planet and reaches the
latitude $20.3^\circ$ S as maximum (no further south), its final
orbit around the Sun will be exactly perpendicular to the ecliptic
plane.
This situation repeats when it ``tries'' to catch the planet and reaches
the latitude $20.3^\circ$ N as maximum (no further north).
The maximum speed given by gravity assist is obtained when the
spacecraft ``chases" the planet on the ecliptic plane and the minimum
when they encounter each other on the ecliptic plane.

Let us check the agreement between our results and the actual
Ulysses's orbit data.
The Ulysses's elevation angle is around $80^\circ$, then
from Table \ref{tab:elevation}, $v = 7.4$ km/s,
which implies that its semi-major axis is about 3.10 AU
(we have used Eq.~(\ref{major}) with
$E/m = v^2/2 - G M_S/a_J$).
This conclusion is quite close with the actual Ulysses's
semi-major axis 3.37 AU (computed by using the first aphelion 5.40 AU
and the first perihelion 1.34 AU, obtained from the web page).
\cite{JPL}
This comparison shows that our 3D slingshot formulation is numerically
acceptable in spite of our assumptions and approximations.
Moreover, it provides physical insight into how
the gravity assist is used to deviate the spacecraft not only on the
ecliptic plane but also away from it, according to the navigator's
desire.

Finally, let us numerically justify our approximation in
Sec.~\ref{toy} for Jupiter case.
The Sun's gravitational force around Jupiter is stronger than that
of Jupiter when objects are located at distances of
$r \gtrsim \sqrt{M_J/M_S}\, a_J = 0.03\, a_J = 334\, R_J$, so that
our claim in Sec.~\ref{toy} is fairly good.


\begin{acknowledgments}
I wish to thank my professor Daniel Azofeifa for getting me
involved in the slingshot effect when I was a graduate student.
I should also thank him for all discussions and comments, now
that I have finally written down my thoughts.
\end{acknowledgments}

\end{document}